\documentclass[review]{article}
\usepackage{lineno, graphicx, latexsym}
\usepackage[font=small,labelfont=bf]{caption}

\newtheorem{thm}{Theorem}
\newtheorem{defin}{Definition}
\newtheorem{lemma}{Lemma}
\newtheorem{prop}{Proposition}
\newtheorem{fact}{Fact}
\def\myqed{\mbox{\hspace{\fill}}\hfill $\Box$}

\title{On approximating tree spanners that are breadth first search trees}

\author{
Ioannis Papoutsakis \\
Kastelli Pediados, Heraklion, Crete, Greece, 700 06
}

\begin{document}
\maketitle

\begin{abstract}
A tree $t$-spanner $T$ of a graph $G$ is a spanning tree of $G$ such that the distance in $T$ between every
pair of verices is at most $t$ times the distance in $G$ between them. There are efficient algorithms that
find a tree $t\cdot O(\log n)$-spanner of a graph $G$, when $G$ admits a tree $t$-spanner. In this paper,
the search space is narrowed to $v$-concentrated spanning trees, a simple family that
includes all the breadth first search trees starting from vertex $v$. In this case, it is not easy to
find approximate tree spanners within factor almost $o(\log n)$. Specifically, let $m$ and $t$ be integers, such
that $m>0$ and $t\geq 7$. If there is an efficient algorithm that receives as input a graph $G$ and a vertex $v$ and
returns a $v$-concentrated tree $t\cdot o((\log n)^{m/(m+1)})$-spanner of $G$, when $G$ admits a
$v$-concentrated tree $t$-spanner, then there is an algorithm that decides 3-SAT in quasi-polynomial time.
\end{abstract}

\section{Introduction}

A tree $t$-spanner $T$ of a graph $G$ is a spanning tree of $G$ such that the distance between every pair of vertices in $T$ is
at most $t$ times the distance between them in $G$.
There are applications of spanners in a variety of areas, such as
distributed computing \cite{Awerbuch85,Peleg89},
communication networks \cite{PelUpf88,PelegReshef}, motion planning and
robotics \cite{Arikati96,Chew89}, phylogenetic analysis
\cite{Bandelt86}, and in embedding finite
metric spaces in graphs approximately \cite{Rabinov98}.
In \cite{Pettie-Low-Dist-Span} it is mentioned that spanners have applications
in approximation algorithms for geometric spaces \cite{Narasimhanbook},
various approximation algorithms \cite{Fakcharoenphol}
and solving diagonally dominant linear systems \cite{Spielman}.

On one hand, in \cite{Bondy89,CaiCor95a} an efficient algorithm
to decide tree 2-spanner admissible graphs is presented. On the
other hand, in \cite{CaiCor95a} it is proved that for
each $t\geq 4$ the problem to decide graphs that admit a tree $t$-spanner
is an NP-complete problem. The complexity status of the tree 3-spanner
problem is unresolved.

There are NP-completeness results for the tree $t$-spanner problem for families of graphs.
In \cite{Fekete01}, it is shown that it is NP-hard to determine the minimum $t$ for which a
planar graph admits a tree $t$-spanner. For any $t\geq4$, the tree $t$-spanner problem is NP-complete
on chordal graphs of diameter at most $t+1$, when $t$ is even, and of diameter at most $t+2$, when 
$t$ is odd \cite{Brandstadtchordal}; note that this refers to the diameter of the graph not to the diameter
of the spanner.  In \cite{Treespannersofsmalldiameter} (which is based on a chapter of \cite{PhDthesis})
it is shown that the problem to determine whether a graph admits a tree $t$-spanner of
diameter at most $t+1$ is tractable, when $t\leq 3$, while it is an
NP-complete problem, when $t\geq 4$. The reduction in this last NP-completeness proof
is used as a building block for the reduction in this article (see subsection~\ref{ss:small_diameter}).

In \cite{Fekete01}, for every $t$, an efficient algorithm to
determine whether a planar graph with bounded face length admits a tree $t$-spanner is
presented. Using a theorem of Logic, the existence of an efficient algorithm to decide
bounded degree graphs that admit a tree $t$-spanner appears in \cite{Fomin}.
Also, for every $t$, an efficient dynamic programming algorithm to
decide tree $t$-spanner admissibility of bounded degree graphs appears in
\cite{Treespannersofboundeddegree}.

The first non trivial approximation algorithm appears in \cite{pelegemek}.
There, an efficient algorithm that finds a tree $t\cdot O(\log n)$-spanner,
when the input graph admits a tree $t$-spanner, is presented. In \cite{DraganK} a different
efficient algorithm achieving similar approximation ratio is presented, using chordal graphs;
it is also given a necessary condition for a graph to admit a tree $t$-spanner.

An alternative definition of the problem of deciding tree $t$-spanner admissible graphs is the following. Let $T$ be
a spanning tree of a graph $G$. The {\em stretch} of a pair of vertices $u$, $v\in G$ is the ratio
of the distance between them in $T$ to the distance between them in $G$. The maximum stretch of
$T$ is the maximum stretch over all pairs of vertices of $G$. The Minimum Max-stretch spanning
Tree problem (MMST) is finding a spanning tree of minimum maximum stretch; i.e. finding a tree $t$-spanner
of a given unweighted graph $G$, such that $G$ does not admit a tree $(t-1)$-spanner. In \cite{PelegReshef} it is
proved that approximating the MMST problem within a factor better than $\frac{1 + \sqrt{5}}{2}$ is NP-hard; note that
this holds for big values of minimum maximum stretch. In \cite{pelegemek}, it is also shown that, for sufficiently big
$t$, it is hard to find a tree $(t+ o(n))$-spanner of a given graph $G$, when $G$ admits a tree $t$-spanner; note
that in this case the minimum maximum stretch is approximated additively. 

An approximation algorithm has to find a good enough spanning tree of the input graph. In this article,
the search space is restricted to $v$-concentrated spanning trees of the input graph $G$, where
$v\in G$ (see definition~\ref{d:concentrated}). The family of $v$-concentrated spanning trees of a graph $G$
is simple, easy to decide, and contains all the breadth first search spanning trees of $G$ with single source
vertex $v$. In this case it is not easy to find approximate tree spanners within factor almost $o(\log n)$.
Specifically, let $m$ and $t$ be integers, such that $m>0$ and $t\geq 7$. Unless there is a
quasi-polynomial time algorithm for 3-SAT, there is no efficient algorithm that receives as input a graph $G$ and a
vertex $v$ and returns a $v$-concentrated tree $t\cdot o((\log n)^{m/(m+1)})$-spanner of $G$, when $G$ admits a
$v$-concentrated tree $t$-spanner (theorem~\ref{t:inapprox}).

\section{Definitions and lemmas}
\label{s:definitions}
In general, terminology of \cite{West} is used. If $G$ is a graph, then $V(G)$ is its {\em vertex set} and
$E(G)$ its {\em edge set}. An edge between vertices $u,v\in G$ is denoted as $uv$.
If $H$ is a subgraph of $G$, then $G[H]$ is the subgraph of $G$ {\em induced} by the vertices
of $H$, i.e. $G[H]$ contains exactly all the vertices of $H$ and all the edges of $G$ between vertices of $H$.

Let $v$ be a vertex of $G$, then $N_G(v)$ is the set of $G$ neighbors of $v$, while
$N_G[v]$ is $N_G(v)\cup \{v\}$; in this paper we consider graphs without loop edges.
The $G$ {\em distance} between two vertices $u,v$ of a connected graph $G$, denoted as $d_G(u,v)$, is the length
of a $u, v$ shortest path in $G$. The $G$ distance between a subgraph $X$ of $G$ and a vertex $v$ of $G$ is
$\min_{x\in X}d_G(x,v)$ and it is denoted as $d_G(X,v)$.
Finally, the $i$th {\em neighborhood} of a vertex $v$ of a graph $G$ is
defined as $N_G^i[v]=\{x\in V(G): d_G(v,x)\leq i\}$. The definition of a tree $t$-spanner follows.

\begin{defin}
A graph $T$ is a tree $t$-spanner of a graph $G$ if and only if $T$ is a
subgraph of $G$ that is a tree and, for every pair $u$ and $v$ of vertices of
$G$, if $u$ and $v$ are at distance $d$ from each other in $G$, then $u$
and $v$ are at distance at most $t\cdot d$ from each other in $T$.
\end{defin}

Note that in order to check whether a spanning tree of a graph $G$
is a tree $t$-spanner of $G$, it suffices to examine pairs of adjacent
in $G$ vertices.

To apply the technique introduced in this article, the search space of spanning trees (towards
finding a tree $t$-spanner) must be narrowed. It seems that the broadest family of spanning
trees this technique can capture is the following.

\begin{defin}
Let $G$ be a graph and $v$ one of its vertices. A spanning tree $T$ of $G$ is $v$-concentrated if and only if
for every $i$, $T[N_G^i[v]]$ is a connected graph.
\label{d:concentrated}
\end{defin}

Clearly, a breadth first search spanning tree of a graph starting from a vertex $v$ is $v$-concentrated.
Also, there can be many\footnote{For example clique $K_n$ has only one breadth first search tree starting from a
vertex $v$ of $K_n$ but it has super-polynomially on $n$ many $v$-concentrated spanning trees.}
$v$-concentrated spanning trees that are not breadth first search spanning trees starting from $v$.
Moreover, note that one can prove\footnote{A $T$ path joining  the endpoints of an edge of $G$ between a vertex
at distance $d$ ($d>0$) from $v$ and a vertex at distance $d-1$ from $v$ can stretch up to $G$ distance
$d'$ away from $v$ before returning back.} the following:
\begin{prop}
\label{p:concentrated}
Let $G$ be a graph that admits a tree $t$-spanner $T$, where $t\geq 1$.
For every vertex $v\in G$ and for every $d\geq0$, the verices in
$N_G^d[v]$ are in the same component of $T[N_G^{d'}[v]]$, where $d'=d+\lfloor\frac{t-1}{2}\rfloor$.
\end{prop}
This proposition hints that every tree $t$-spanner is loosely ``concentrated" around each vertex $v$.

An instance of 3-SAT is a set of clauses, where each clause is the disjunction of exactly 3 distinct literals; a literal
is a boolean variable or its negation. The 3-SAT problem is to decide whether there is a truth assignment to the
variables of a given instance, such that all its clauses are satisfied. Note that if a clause contains less than 3
variables, then both a variable and its negation appear in the clause; so, the clause is satisfied by every truth
assignment. Therefore, it suffices to examine instances for which each clause contains exactly 3 variables.
In this article, it is assumed that each clause of an instance of 3-SAT contains
exactly 3 distinct variables.

Let $f$ and $g$ be functions from the set of graphs to the set of non negative integers. Then, $f$ is $O(g)$ if
and only if there exist graph $G_0$ and integer $C$, such that $f(G)\leq Cg(G)$, for every graph $G$ with $|V(G)|>|V(G_0)|$.
Also, $f$ is $o(g)$ if and only if for every $\epsilon>0$ there is a graph $G_{\epsilon}$ such
that $f(G)<\epsilon g(G)$ for every graph $G$ with $|V(G)|>|V(G_{\epsilon})|$.

To define the running time of an algorithm, assume that the algorithm is implemented by a deterministic Turing machine.
For this, objects, such as instances of problems or outputs of algorithms, are encoded as 0-1 strings. For example,
instances of 3-SAT can be encoded as 0-1 strings; then, the size of an instance of 3-SAT is the length of its encoding. 
An algorithm runs in time $f(n)$ if there is a deterministic Turing machine M that implements the algorithm and the time required by
M on each input of length $n$ is at most $f(n)$. If an algorithm runs in polynomial time, then the algorithm is called {\em efficient}.

\section{Description of the reduction}
Algorithm {\tt reduction} is presented in figure~\ref{f:reduction}; it takes as input an instance $\phi$
of 3-SAT and an integer $h>1$, while it returns a graph $G$. Here, $h$ is a parameter set in the
proof of theorem~\ref{t:inapprox} and
depends on the number of variables of $\phi$; its choice is crucial for relating the finding of a not too bad
approximate tree spanner of $G$ to a low enough running time for deciding satisfiability of $\phi$ upon such a tree spanner.
Given $\phi$, graphs are constructed by calling function {\tt get\_bb} in figure~\ref{f:building},
which become the building blocks of the final graph $G$. These building blocks are put together in a tree like structure of height $h$.

\subsection{Relation to a known NP-complete problem}
\label{ss:small_diameter}
In \cite{Treespannersofsmalldiameter} it is proved that it is an NP-complete problem to decide whether a graph
admits a tree $t$-spanner of diameter at most $t+1$, for $t\geq 4$. The reduction there is from 3-SAT. It turns
out that for $t=7$, graphs being built for the sake of this NP-completeness reduction can be stacked one on top of the other
like building blocks. Then, a final graph $G$ is constructed by stacking building blocks, starting with a path
having a central vertex $v$. This way,
the difficulty of finding a tree 7-spanner locally propagates, creating a chasm;  in any easily\footnote{Meaning a tree
spanner that does not solve the difficult tree 7-spanner problem locally.} found tree spanner of
$G$ that is also concentrated around $v$, some two vertices high in a stack are adjacent
in $G$ but far apart in the tree spanner.

Note that in \cite{Treespannersofsmalldiameter} it is essential to prove the fact that if a graph $G$ admits a tree
$t$-spanner of diameter at most $t+1$, then $G$ admits a tree $t$-spanner that is a breadth first search tree.
For bigger diameters, this fact does not hold; so, the search space for tree spanners must somehow be narrowed to
spanning trees that are concentrated around a central vertex.

\subsection{Formation of building block}
\label{s:building}
\begin{figure}[htbp]
\begin{center}
\includegraphics[width=.7\textwidth]{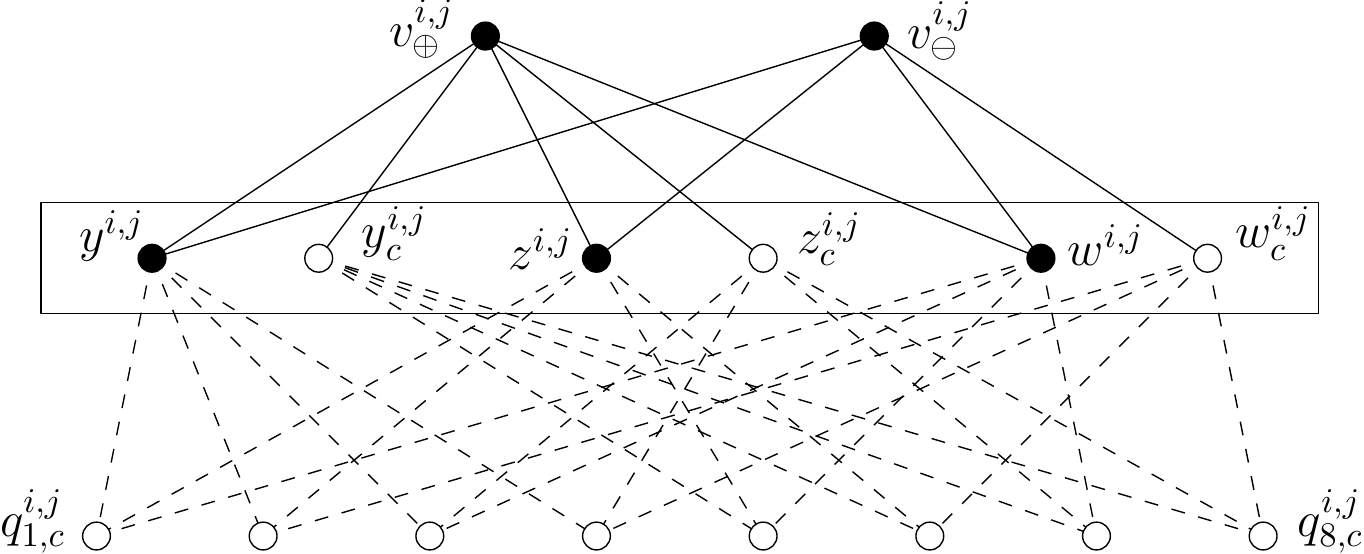}
\caption{Part of graph $G^{i,j}$, being the output of function {\tt get\_bb}. Let $c$ be clause $(y\vee z\vee\neg w)$.
Vertices related to clause $c$ plus the distinct vertices $v_{\oplus}^{i,j}$ and $v_{\ominus}^{i,j}$ are shown. Vertices in the
rectangle are the vertices in array $g$. The dashed edges are these determined by matrix $M$. All the edges of $G^{i,j}$
incident to the white vertices are shown in the figure.}
\label{f:partofbuilding}
\end{center}
\end{figure}

Function {\tt get\_bb} in figure~\ref{f:building} receives as input an instance $\phi$ of 3-SAT and
two integers $i$, $j$ and constructs a graph.
Integers $i$, $j$ become labels of vertices of the output graph in order to distinguish them among copies of this graph;
also, the output graph is denoted as $G^{i,j}$ by the main function that calls {\tt get\_bb}.  A part of graph $G^{i,j}$ is
shown in figure~\ref{f:partofbuilding}.

\begin{figure}[htbp]
\begin{center}
\begin{tt}
\fbox{
\parbox{0.93\linewidth}{
{\bf Function get\_bb}($\phi$,$i$,$j$)\newline
{\bf Input}. A nonempty instance $\phi$ of 3SAT and two integers $i$, $j$.\newline

$V^{i,j}=\{v_{\oplus}^{i,j}, v_{\ominus}^{i,j}\}$\newline
{\bf for} (variable $x$ of $\phi$) $V^{i,j}=V^{i,j}\cup\{x^{i,j}\}$\newline
{\bf for} (clause $c$ in $\phi$)\{\newline
\mbox{\hspace{.5cm}}{\bf for} (variable $x$ of $c$) $V^{i,j}=V^{i,j}\cup\{x_c^{i,j}\}$\newline
\mbox{\hspace{.5cm}}{\bf for} ($r=1$ to 8) $V^{i,j}=V^{i,j}\cup\{q_{r,c}^{i,j}\}$\newline
\}\newline
$E^{i,j}=\emptyset$\newline
{\bf for} (variable $x$ of $\phi$) $E^{i,j}=E^{i,j}\cup\{x^{i,j}v_{\oplus}^{i,j},x^{i,j}v_{\ominus}^{i,j}\}$\newline
{\bf for} (clause $c$ in $\phi$)\{\newline
\mbox{\hspace{.5cm}}{\bf Let} $y$,$z$, and $w$ be the variables of $c$\newline
\mbox{\hspace{.5cm}}$g=[y^{i,j},y_c^{i,j},z^{i,j},z_c^{i,j},w^{i,j},w_c^{i,j}]$\newline
\mbox{\hspace{.5cm}}{\bf for} ($k=1$ to 6) {\bf for} ($l=1$ to 8)\newline 
\mbox{\hspace{1cm}}{\bf if} ($M[k,l]=1$) $E^{i,j}=E^{i,j}\cup\{g[k]q_{l,c}^{i,j}\}$\hfill /*(1)\newline
\mbox{\hspace{.5cm}}{\bf for} ($x$ in $\{y,z,w\}$)\newline
\mbox{\hspace{1cm}}{\bf if} ($x$ appears positive in $c$) $E^{i,j}=E^{i,j}\cup\{x_c^{i,j}v_{\oplus}^{i,j}\}$\newline
\mbox{\hspace{1cm}}{\bf else} $E^{i,j}=E^{i,j}\cup\{x_c^{i,j}v_{\ominus}^{i,j}\}$\newline
\}\newline
{\bf return} ($V^{i,j}, E^{i,j}$)
}}
\end{tt}
\caption{Function {\tt get\_bb($\phi$,$i$,$j$)} that forms a graph given an instance $\phi$ of 3-SAT, which becomes a building
block of  the final graph. Parameters $i$, $j$ are used to label each building block. Matrix $M$ used in line (1) is defined outside
of the function and is presented in section~\ref{s:building} (equation~\ref{e:matrix_M}). Elements in array $g$ and matrix
$M$ are numbered starting from 1, not 0. Note that it is essential, as pointed out in section~\ref{s:definitions}, that
each clause in $\phi$ contains exactly 3 variables.}
\label{f:building}
\end{center}
\end{figure}

The vertex set of $G^{i,j}$ is generated. First, two distinct vertices $v_{\oplus}^{i,j}$ and $v_{\ominus}^{i,j}$ are
placed into the vertex set of $G^{i,j}$; these vertices will be used by the {\tt reduction} algorithm to glue the
new building block to the existing construction. Also, $v_{\oplus}^{i,j}$ will ``attract" the positive standings of variables
in clauses (similarly, $v_{\ominus}^{i,j}$ the negative). Second, each
Boolean variable of $\phi$ gives rise to a vertex of $G^{i,j}$; for each variable $x$ of $\phi$ vertex $x^{i,j}$ of
$G^{i,j}$ is generated.

Third, for each clause $c$ in $\phi$, 11 new
vertices of $G^{i,j}$ are generated. Specifically, 3 vertices are for the presence of each of the 3 variables in $c$ and are
distinct from the vertices generated for the variables of $\phi$; these vertices carry the subscript $c$. For example,
if $c$ contains variables $y$, $z$, and $w$, then $G^{i,j}$ contains vertices $y_c^{i,j}$, $z_c^{i,j}$, and $w_c^{i,j}$.
Additionally, the remaining 8 vertices take letter $q$, are numbered from 1 to 8, and carry the subscript $c$ as well;
i.e. $G^{i,j}$ contains vertices $q_{1,c}^{i,j}, q_{2,c}^{i,j},\ldots, q_{8,c}^{i,j}$.

Then, the edges of $G^{i,j}$ are formed. First the vertex that corresponds to each variable of $\phi$
becomes adjacent to both of the distinct vertices $v_{\oplus}^{i,j}$ and $v_{\ominus}^{i,j}$.
Second, for each clause $c$ in $\phi$, edges
between the 14 vertices related to $c$ are placed; 11 vertices have been generated for clause $c$ and 3 vertices
correspond to variables of $\phi$ that participate in $c$. These 14 vertices are partitioned in two groups; the one group
contains all 8 vertices denoted with letter $q$ and the other group the remaining 6 vertices. Each vertex in the group of
8 (the $q$ vertices) took a number when it was created; thusly, the vertices of this group are numbered from 1 to 8.
The vertices in the group of 6 are placed in an array $g$ and in this way are numbered from 1 to 6 (figure~\ref{f:building});
for example, if $c$ contains variables $y$, $z$, and $w$, then $g=[y^{i,j},y_c^{i,j},z^{i,j},z_c^{i,j},w^{i,j},w_c^{i,j}]$.
Having numbered the vertices within each group, the adjacencies between
the two groups are determined by the following matrix:
\begin{equation}
\label{e:matrix_M}
M=\left[ \begin{array}{cccccccc}
       1 & 1 & 1 & 1 & 0 & 0 & 0 & 0 \\
       0 & 0 & 0 & 0 & 1 & 1 & 1 & 1 \\
       1 & 1 & 0 & 0 & 1 & 1 & 0 & 0 \\
       0 & 0 & 1 & 1 & 0 & 0 & 1 & 1 \\
       1 & 0 & 1 & 0 & 1 & 0 & 1 & 0 \\
       0 & 1 & 0 & 1 & 0 & 1 & 0 & 1 \end{array} \right]
\end{equation}

Matrix $M$ has two main properties. It consists of three pairs of complementary to each other rows (for example, the
first row is the complement of the second). Also, if a sub-matrix of $M$ consisting of whole rows of $M$ contains at least one
1 in each column, then the sub-matrix must contain at least one pair of complementary to each other rows.

Third, there are a few more edges incident to vertices related to $c$. To indicate the standing (negation or not) of each variable $x$
of $c$, vertex $x_c^{i,j}$ is made adjacent to either $v_{\ominus}^{i,j}$ or $v_{\oplus}^{i,j}$. Note that vertex $x_c^{i,j}$ is related only to clause $c$; in contrast, vertex $x^{i,j}$ may be related to many clauses.

\subsection{Putting building blocks together}

\begin{figure}[htbp]
\begin{center}
\begin{tt}
\fbox{
\parbox{0.9\linewidth}{
{\bf Algorithm reduction}($\phi$,$h$)\newline
{\bf Input}. A nonempty instance $\phi$ of 3-SAT and an integer $h>1$.\newline

{\bf Let} $G$ be the path $q_1,p_1,v,p_2,q_2$\newline
$G^{1,1}=$ {\bf get\_bb}($\phi,1,1$)\newline
$G=G\cup G^{1,1}$\newline
{\bf identify} $q_1$ with $v_{\oplus}^{1,1}$; {\bf identify} $q_2$ with $v_{\ominus}^{1,1}$\newline
$b[1]=1$; $j=0$\newline
{\bf for} ($i=2$ to $h$)\{\newline
\mbox{\hspace{.5cm}}{\bf for} ($s=1$ to $b[i-1]$)\newline
\mbox{\hspace{1cm}}{\bf for} (clause $c$ in $\phi$)\newline
\mbox{\hspace{1.5cm}}{\bf for} ($r=1$ to $7$)\{\newline
\mbox{\hspace{2cm}}$j=j+1$\newline
\mbox{\hspace{2cm}}$G^{i,j}=$ {\bf get\_bb}($\phi,i,j$)\newline
\mbox{\hspace{2cm}}$G=G\cup G^{i,j}$\newline
\mbox{\hspace{2cm}}{\bf identify} $q_{r,c}^{i-1,s}$ with $v_{\oplus}^{i,j}$; {\bf identify} $q_{r+1,c}^{i-1,s}$ with $v_{\ominus}^{i,j}$\newline
\mbox{\hspace{1.5cm}}\}\newline
\mbox{\hspace{.5cm}}$b[i]=j$; $j=0$;\newline
\}\newline
{\bf return} $G$
}}
\end{tt}
\caption{Algorithm {\tt reduction($\phi$,$h$)} that forms a graph given an instance $\phi$ of 3-SAT; building
blocks formed by function {\tt get\_bb} in figure~\ref{f:building} are put together in a tree like structure of hight $h$.
Variable $b[i]$ stores the number of building blocks in layer $i$, while variable $s$ iterates over the
building blocks of layer $i-1$. Finally, variable $r$ is used to iterate over consecutive pairs of $q$ vertices related
to clause $c$ in building block $G^{i-1,s}$.}
\label{f:reduction}
\end{center}
\end{figure}

The construction of the final graph $G$ starts with a path of length 4, having $v$ as its central vertex.
Function {\tt get\_bb} on input ($\phi$, $i$, $j$) provides building block $G^{i,j}$, where $i$ and $j$ are the
indexes of the block. Then, building blocks with various indexes are added in layers to the existing structure.
The first index of a building block indicates the layer that the block is placed in; i.e. {\em layer} $i$ of $G$ contains
exactly all graphs produced by calling {\tt get\_bb} on input ($\phi$, $i$, $j$) for various $j$, while
executing algorithm {\tt reduction} on input ($\phi$,$h$).

\begin{figure}[htbp]
\begin{center}
\includegraphics[width=.9\textwidth]{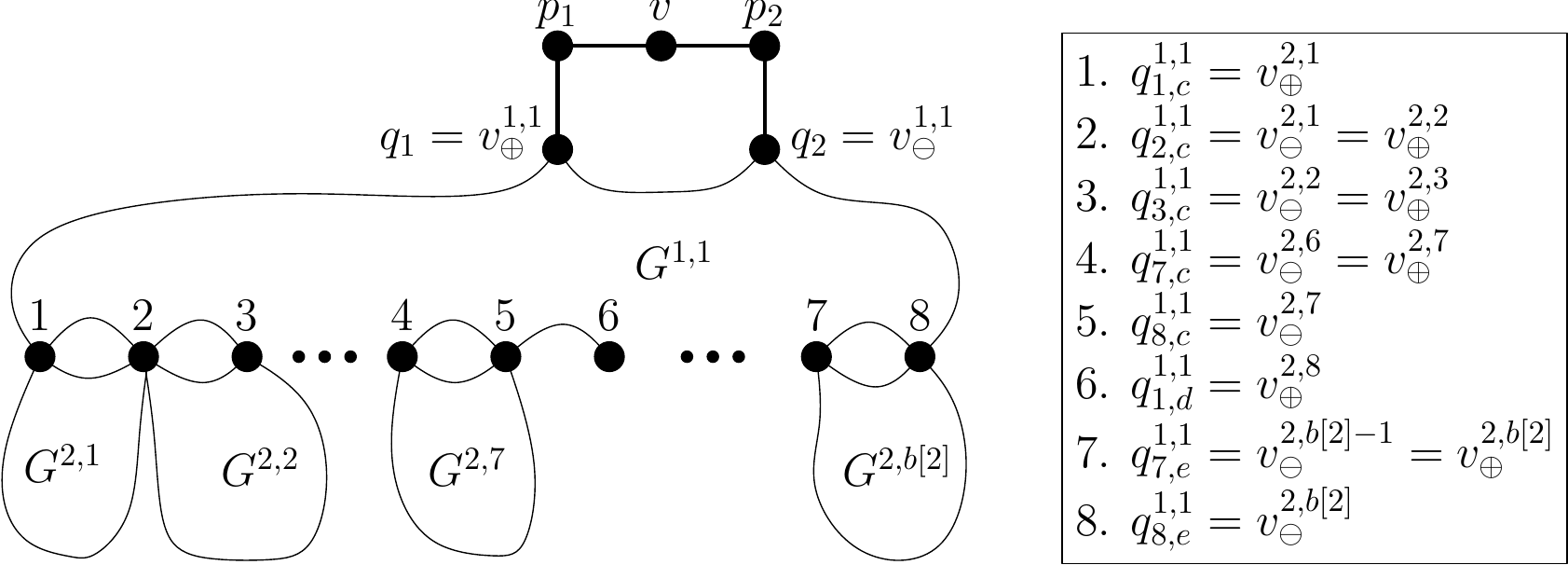}
\caption{Part of output graph $G$ of algorithm {\tt reduction} in figure~\ref{f:reduction} for $h=2$. Only three clauses are
involved here, namely $c$, $d$, and $e$, where $c$ is examined first in the {\tt for} loops, $d$ second, and $e$ last. The
table on the right hand side shows the vertices of $G$ that correspond to the numbered vertices in the figure; for example
number 2 corresponds to 3 vertices of 3 different building blocks, which all have been identified to one vertex of $G$. Note that
there is no building block attached to pair 5, 6 of $q$ vertices of $G^{1,1}$, since these are related to different clauses.
The number of building blocks in layer 2 is $b[2]$.}
\label{f:final}
\end{center}
\end{figure}

The first building block, graph $G^{1,1}$, is attached to the endpoints of the primary path by identifying the one
endpoint of the path with vertex $v_{\oplus}^{1,1}$ and the other endpoint with vertex $v_{\ominus}^{1,1}$. Layer 1
contains only one graph, namely $G^{1,1}$. After this first layer (command {\tt for ($i=2$ to $h$)} in figure~\ref{f:reduction}),
at each step of the construction, a new building block is attached to a pair of consecutive $q$ vertices of
the previous layer. Specifically, building block $G^{i,j}$ is attached to building block $G^{i-1,s}$ by identifying 
$q_{r,c}^{i-1,s}$ with $v_{\oplus}^{i,j}$ and $q_{r+1,c}^{i-1,s}$ with $v_{\ominus}^{i,j}$;
in figure~\ref{f:reduction}, $j$ counts the building blocks added to the new layer, $s$ is used to iteratively
examine the building blocks of the previous layer (which are $b[i-1]$ in number), $c$ iterates over the clauses in $\phi$,
and $r$ iterates over the 7 first $q$ vertices related to clause $c$ in building block $G^{i-1,s}$. The process adds iteratively
blocks to layer $i$ until all pairs of consecutive $q$ vertices of layer $i-1$ are covered and then continues with the
next layer. The process halts when $h$ layers are completed. A summary picture of a two layer such graph is shown in figure~\ref{f:final}. 

\section{Hardness of approximation}
\begin{lemma}
For every satisfiable instance of 3-SAT $\phi$ and for every $h>1$, graph $G$ returned by algorithm {\tt reduction} in
figure~\ref{f:reduction} on input $(\phi,h)$ admits a $v$-concentrated tree 7-spanner. 
\label{l:7spanner}
\end{lemma}

{\em Proof}.
Let $a$ be a truth assignment that satisfies $\phi$. Let $T$ be the graph returned by algorithm {\tt tree\_7-spanner}
in figure~\ref{f:7spanner} on input $(G,\phi,a)$.  Part of a building block $G^{i,j}$ of $G$ where edges of $T$ are shown
appears in figure~\ref{f:part7spanner}.

\begin{figure}[htbp]
\begin{center}
\begin{tt}
\fbox{
\parbox{0.9\linewidth}{
{\bf Algorithm tree\_7-spanner}($G$, $\phi$, $a$)\newline
{\bf Input}. A graph $G$, an instance $\phi$ of 3-SAT, and a truth assignment $a$.\newline

$V=V(G)$; $E=\{q_1p_1, p_1v, vp_2, p_2q_2\}$\newline
{\bf for} (building block $G^{i,j}$ of $G$)\{\newline
\mbox{\hspace{.5cm}}{\bf for} (variable $x$ of $\phi$)\newline
\mbox{\hspace{1cm}}{\bf if} ($a(x)=1$) $E=E\cup\{x^{i,j}v_{\oplus}^{i,j}\}$\newline
\mbox{\hspace{1cm}}{\bf else} $E=E\cup\{x^{i,j}v_{\ominus}^{i,j}\}$\newline
\mbox{\hspace{.5cm}}{\bf for} (clause $c$ in $\phi$)\{\newline
\mbox{\hspace{1cm}}{\bf for} (variable $x$ of $c$)\newline
\mbox{\hspace{1.5cm}}$E=E\cup E(G[\{v_{\oplus}^{i,j}, v_{\ominus}^{i,j}, x^{i,j}_c\}])$\hfill /*(1)\newline
\mbox{\hspace{1cm}}{\bf Let} $z$ be a variable of $c$ that makes $c$ true through $a$\newline
\mbox{\hspace{1cm}}$Q=\bigcup_{r=1}^{r=8}\{q^{i,j}_{r,c}\}$\newline
\mbox{\hspace{1cm}}$E=E\cup E(G[Q\cup\{z^{i,j},z^{i,j}_c\}])$\hfill /*(2)\newline
\mbox{\hspace{.5cm}}\}\newline
\}\newline
{\bf return} ($V, E$)
}}
\end{tt}
\caption{Algorithm {\tt tree\_7-spanner($G$, $\phi$, $a$)} that constructs a tree 7-spanner of $G$, when $G$ is
the output of algorithm {\tt reduction($\phi$,$h$)} in figure~\ref{f:reduction}, where $\phi$ is a satisfiable instance of
3-SAT; also, $a$ is a truth assignment that satisfies $\phi$. It is assumed that the building blocks of $G$ are known to
this algorithm, as part of the input graph $G$. As described in the proof
of lemma~\ref{l:7spanner}, the command in line (1) results in adding one edge to $E$, while the command in line (2)
results in adding 8 edges to $E$.}
\label{f:7spanner}
\end{center}
\end{figure}

An essential fact hinting that the tree 7-spanner problem is solved locally is proved first. Let $G^{i,j}$ be a building
block of $G$ and $c$ a clause in $\phi$. Also, let $Q_c^{i,j}=\bigcup_{r=1}^{r=8}\{q_{r,c}^{i,j}\}$. 
There is one variable $z$ of $c$ which is used to form $T$ edges incident to vertices in $Q_c^{i,j}$ within $G^{i,j}$ (see line (2)
in figure~\ref{f:7spanner}). Here, $z^{i,j}$ and $z_c^{i,j}$ correspond to complementary to each other rows of
matrix $M$; so, each vertex in $Q_c^{i,j}$ is adjacent to one of $z^{i,j}$ or $z_c^{i,j}$. But $z$  is a variable that
makes $c$ true through $a$; so, both of $z^{i,j}$ and $z_c^{i,j}$ are adjacent in $T$ to the same
vertex $v_{\oplus}^{i,j}$ or $v_{\ominus}^{i,j}$ (see figure~\ref{f:part7spanner}). Therefore,
\begin{fact}
For every building block $G^{i,j}$ of $G$ and for every clause $c$ in $\phi$, the $T$ distance between any two
vertices in $Q_c^{i,j}$ is at most 4.
\label{fact:local}
\end{fact}

Clearly, $T$ spans $V(G)$. Let $P$ be path
$q_1,p_1,v,p_2,q_2$. Let $b(i')$ be the number of building blocks in layer $i'$ of $G$. For every $i$, $0\leq i\leq h$,
let $V^i=V(P)\cup\bigcup_{i'=1}^{i'=i}\bigcup_{j=1}^{j=b(i')}V(G^{i',j})$; then, let $G^i=G[V^i]$. To picture $G^i$,
it is the subgraph of $G$ induced by the first $i$ layers of $G$ plus path $P$.
Clearly, $G^0=P$ and $G^h=G$. It is proved by induction
on $i$ that $T[G^i]$ is a $v$-concentrated tree 7-spanner of $G^i$. For the base case, $i=0$, both of $T[G^0]$ and
$G^0$ are equal to path $P$, which has $v$ as its central vertex.

Consider a building block $G^{i,j}$ in layer $i$, where $1\leq i\leq h$. Let $X$ be the set of variables of $\phi$; also, let
$X^{i,j}=\bigcup_{x\in X}\{x^{i,j}\}$. Then, for every $x\in X$, vertex $x^{i,j}$ is adjacent in $T$ to either
$v_{\oplus}^{i,j}$ or
$v_{\ominus}^{i,j}$ depending on the value given by $a$ for $x$. Therefore, $T[X^{i,j}\cup\{v_{\oplus}^{i,j}, v_{\ominus}^{i,j}\}]$ consists of exactly
two trees, one containing $v_{\oplus}^{i,j}$ and the other $v_{\ominus}^{i,j}$. Note that both of edges
$x^{i,j}v_{\oplus}^{i,j}$ and $x^{i,j}v_{\ominus}^{i,j}$ are edges of $G$; so, $T[X^{i,j}\cup\{v_{\oplus}^{i,j}, v_{\ominus}^{i,j}\}]$ is a subgraph of $G$. Towards proving that $T$ is $v$-concentrated, observe that
each vertex $u$ in $X^{i,j}$ is adjacent in $T$ to a vertex (namely $v_{\oplus}^{i,j}$ or
$v_{\ominus}^{i,j}$), which is closer than $u$ to $v$ in $G$.

\begin{figure}[htbp]
\begin{center}
\includegraphics[width=.7\textwidth]{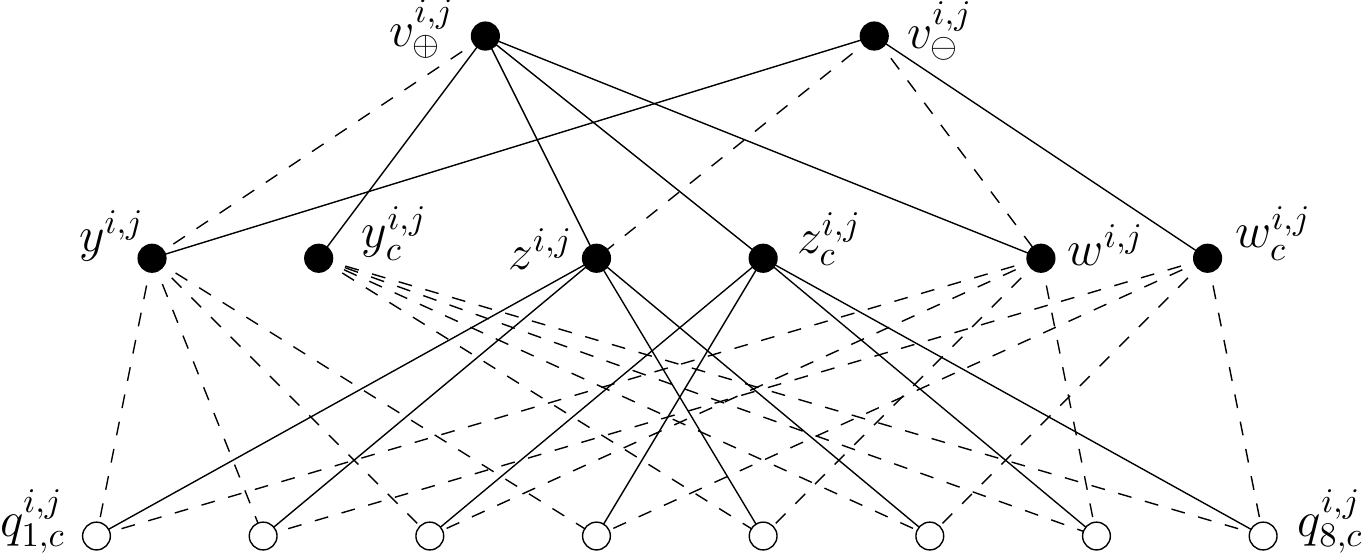}
\caption{Part of graph $G^{i,j}$, being a building block of graph $G$ returned by algorithm {\tt reduction} in
figure~\ref{f:reduction} on input $(\phi,h)$, where $\phi$ is a satisfiable instance of 3-SAT and $h>1$.
Let $c$ be clause $(y\vee z\vee\neg w)$ of $\phi$.
Vertices related to clause $c$ plus the distinct vertices $v_{\oplus}^{i,j}$ and $v_{\ominus}^{i,j}$ are shown. Here,
$a(y)=0$, $a(z)=1$, and $a(w)=1$, where $a$ is a truth assignment that satisfies $\phi$.
Solid edges belong to $T$, which is the tree returned by algorithm {\tt tree\_7-spanner}
in figure~\ref{f:7spanner} on input $(G,\phi,a)$; note that $z$ is the only variable that  makes $c$ true through $a$.
Observe that the $T$ distance between a pair of white vertices (the $q$ vertices) is at most 4.}
\label{f:part7spanner}
\end{center}
\end{figure}

Let $C$ be the set of clauses of $\phi$. Also, for every clause $c\in C$, let $X_c$ be the set containing the three variables
of $c$. Let $X_c^{i,j}=\bigcup_{x\in X_c}\{x_c^{i,j}\}$. Then (see line (1) in figure~\ref{f:7spanner}),
every vertex in $X_c^{i,j}$ is adjacent in $T$ to either
$v_{\oplus}^{i,j}$ or $v_{\ominus}^{i,j}$ with the one and only edge\footnote{Let $x$ be a variable that appears in $c$.
Then, vertex $x_c^{i,j}$ is adjacent in $G$ to either $v_{\oplus}^{i,j}$ or $v_{\ominus}^{i,j}$, depending on the standing
(negation or not) of $x$ in $c$ (see figure~\ref{f:building}). Also, $v_{\oplus}^{i,j}$ is not adjacent to $v_{\ominus}^{i,j}$
in $G$. Therefore, $E(G[\{v_{\oplus}^{i,j}, v_{\ominus}^{i,j}, x^{i,j}_c\}])$ contains only one edge.} of $G$ between these 3
vertices. Therefore, $T[X^{i,j}\cup\{v_{\oplus}^{i,j}, v_{\ominus}^{i,j}\}\cup\bigcup_{c\in C}X_c^{i,j}]$ consists of exactly
two subtrees of $G$, one containing $v_{\oplus}^{i,j}$ and the other $v_{\ominus}^{i,j}$. Again, each vertex $u$ in
$\bigcup_{c\in C}X_c^{i,j}$ is adjacent in $T$ to a vertex (namely $v_{\oplus}^{i,j}$ or
$v_{\ominus}^{i,j}$), which is closer than $u$ to $v$ in $G$.

Let $Q_c^{i,j}=\bigcup_{r=1}^{r=8}\{q_{r,c}^{i,j}\}$. There is only one
variable $z$ in $c$ which is used to form $T$ edges incident to vertices in $Q_c^{i,j}$ within $G^{i,j}$ (see line (2)
in figure~\ref{f:7spanner}). But $z^{i,j}$ and $z_c^{i,j}$ correspond to complementary to each other rows of
matrix $M$; so, each vertex in $Q_c^{i,j}$ is adjacent to exactly one of $z^{i,j}$ or $z_c^{i,j}$. Also, note
that $G[Q_c^{i,j}]$ doesn't have any edges and $z^{i,j}$ is not adjacent to $z_c^{i,j}$ in $G$. Therefore,
$T[G^{i,j}]$ consists of exactly two subtrees of $G$, one containing $v_{\oplus}^{i,j}$ and the other $v_{\ominus}^{i,j}$, since
$V(G^{i,j})=X^{i,j}\cup\{v_{\oplus}^{i,j}, v_{\ominus}^{i,j}\}\cup\bigcup_{c\in C}(X_c^{i,j}\cup Q_c^{i,j})$.
Finally, each vertex $u$ in $Q_c^{i,j}$ is adjacent in $T$ to a vertex (namely $z^{i,j}$ or $z_c^{i,j}$), which is closer
than $u$ to $v$ in $G$.

By induction hypothesis, $T[G^{i-1}]$ is a subtree of $G$.
So, graph $T[G^i]$ is formed upon tree $T[G^{i-1}]$ by attaching 0, 1, or 2 subtrees of $G$ to each vertex of $T[G^{i-1}]$.
These attached trees are vertex disjoint to each other, with only exceptions when 2 trees are attached to the same vertex
$u$ and in such cases these 2 trees have only $u$ as a common vertex. Hence, $T[G^i]$ is a subtree of $G$.

Also, $T[G^i]$ is not only $v$-concentrated but a breadth first search tree of $G^i$ starting from $v$ as well; by construction
of $T[G^i]$, every vertex $u$ of $G^i$ (where $u\not= v$) is adjacent in $T[G^i]$ to a vertex closer than $u$ to
$v$ in $G^i$.

If $i=1$, then $G^i$ has only one layer and only one building block; in this case $d_T(v_{\oplus}^{i,j},v_{\ominus}^{i,j})=4$, because of path $P$.
If $i>1$, then $G^{i,j}$ is attached to some building block $G^{i-1,s}$, by identifying pair $v_{\oplus}^{i,j},v_{\ominus}^{i,j}$
with a pair of vertices in $Q_{c'}^{i-1,s}$, where $c'$ is a clause in $\phi$ (similarly to set $Q_c^{i,j}$ above,
$Q_{c'}^{i-1,s}=\bigcup_{r=1}^{r=8}\{q_{r,c'}^{i-1,s}\}$). Conclusively, by fact~\ref{fact:local},
$d_T(v_{\oplus}^{i,j},v_{\ominus}^{i,j})\leq 4$.

By induction hypothesis $T[G^{i-1}]$ is a 7-spanner of $G^{i-1}$.
In order to prove that $T[G^i]$ is a 7-spanner of $G^i$ it suffices to examine $T$ distances between endpoints of
non $T$ edges of $G^{i,j}$. Each vertex in $X^{i,j}$ is at $T$ distance 1 from $v_{\oplus}^{i,j}$ or $v_{\ominus}^{i,j}$,
so the $T$ distance between
endpoints of non $T$ edges of $G^{i,j}[X^{i,j}\cup\{v_{\oplus}^{i,j}, v_{\ominus}^{i,j}\}]$ is at most 5, because 
$d_T(v_{\oplus}^{i,j},v_{\ominus}^{i,j})\leq 4$. It remains to examine non $T$ edges determined by matrix $M$. Each vertex
in $\bigcup_{c\in C}Q_c^{i,j}$ is at $T$ distance 2 from $v_{\oplus}^{i,j}$ or $v_{\ominus}^{i,j}$. Also, each vertex in
$X^{i,j}\cup\bigcup_{c\in C}X_c^{i,j}$ is at $T$ distance 1 from $v_{\oplus}^{i,j}$ or $v_{\ominus}^{i,j}$. Therefore, the 
$T$ distance between a vertex in $\bigcup_{c\in C}Q_c^{i,j}$ and a vertex in $X^{i,j}\cup\bigcup_{c\in C}X_c^{i,j}$
is at most 7, again because $d_T(v_{\oplus}^{i,j},v_{\ominus}^{i,j})\leq 4$.\myqed

\begin{thm}
Let $m$ and $t$ be integers, such that $m>0$ and $t\geq 7$. Also, let $n'$ and $f$ be functions from the set of graphs to the
non negative integers, such that $n'(G)=|V(G)|$, for every graph $G$, and $f$ is $o((\log n')^{\frac{m}{m+1}})$.
If there is an efficient algorithm that receives as input a graph $G$ and a vertex $v$ and returns a $v$-concentrated
tree $t\cdot f(G)$-spanner of $G$, when $G$ admits a $v$-concentrated tree $t$-spanner, then there is an algorithm that
decides 3-SAT in $2^{O((\log n)^{m+1})}$ time.
\label{t:inapprox}
\end{thm}

{\em Proof}.
Since $f$ is $o((\log n')^{\frac{m}{m+1}})$, for every $\epsilon>0$ there is an $H_{\epsilon}$ such
that $f(H)<\epsilon (\log n'(H))^{\frac{m}{m+1}}$ for every $H$ with $|V(H)|>|V(H_{\epsilon})|$.
Let $H'$ be the graph $H_{\epsilon}$ that corresponds to $\epsilon=\frac{4}{t12^{\frac{m}{m+1}}}$.
Let {\tt get\_spanner} be the approximation algorithm assumed by this theorem. Let $\phi$ be a nonempty
instance\footnote{As pointed out in section~\ref{s:definitions}, each clause of an instance of
3-SAT contains exactly 3 variables.} of 3-SAT. It is proved that algorithm {\tt 3-SAT} in figure~\ref{f:3-SAT}
on input $\phi$ returns {\tt YES} if and only if $\phi$ is satisfiable. 

\begin{figure}[htbp]
\begin{center}
\begin{tt}
\fbox{
\parbox{0.9\linewidth}{
{\bf Algorithm 3-SAT}($\phi$)\newline
{\bf Input}. A nonempty instance $\phi$ of 3-SAT.\newline

{\bf Let} $n$ be the number of variables of $\phi$\newline
$G$={\bf reduction}($\phi$, $\lceil(\log n)^m\rceil$)\newline
{\bf if} ($|V(G)|\leq |V(H')|$)\newline
\mbox{\hspace{.5cm}}solve $\phi$ exhaustively and return appropriately\newline 
$T$={\bf get\_spanner}($G, v$)\newline
{\bf for} (building block $G^{i,j}$ of $G$)\{\newline
\mbox{\hspace{.5cm}}{\bf for} (variable $x$ of $\phi$)\newline
\mbox{\hspace{1cm}}{\bf if} ($x^{i,j}v_{\oplus}^{i,j}\in E(T)$) $a(x)=1$\newline
\mbox{\hspace{1cm}}{\bf else} $a(x)=0$\newline
\mbox{\hspace{.5cm}}{\bf if} (truth assignment $a$ satisfies $\phi$)\newline
\mbox{\hspace{1cm}}{\bf return} YES\newline
\}\newline
{\bf return} NO
}}
\end{tt}
\caption{Algorithm {\tt 3-SAT}($\phi$) receives as input a nonempty instance $\phi$ of 3-SAT and decides whether it is
satisfiable. Constant $m$ and graph $H'$ are defined outside of the algorithm; $m$ is a positive integer introduced in
theorem~\ref{t:inapprox}, while $H'$ is given in the first paragraph of its proof.
Algorithm {\tt reduction} is presented in figure~\ref{f:reduction}.
Algorithm {\tt get\_spanner} is not given explicitly but its existence is assumed by the same theorem.
It is assumed that the decomposition of $G$ into building blocks is given too, when $G$ is returned by
algorithm {\tt reduction}.}
\label{f:3-SAT}
\end{center}
\end{figure}

For the necessity, algorithm {\tt 3-SAT} returns {\tt YES} on input $\phi$, only when it finds a truth assignment that
satisfies $\phi$.

For the sufficiency, assume that $\phi$ is satisfiable.
Let $n$ be the number of variables\footnote{There is a way to encode instances of 3-SAT as 0-1 strings,
such that the size of an encoding of an instance $\phi$  is polynomially
bounded by the number of the variables in $\phi$. So, because of the $\log$ in the description of the running time and
the $O$ notation that follows, the size of an instance of 3-SAT can be considered as the number of variables it contains.}
of $\phi$. So, $\phi$ has at most $8n^3$ clauses. Set $h=\lceil(\log n)^m\rceil$. Let $G$ be the
output of algorithm {\tt reduction} in figure~\ref{f:reduction} on input ($\phi$,$h$).
Note that $n\geq 3$, because $\phi$ is nonempty and each of its clauses contains 3 distinct variables;
so, $h=\lceil(\log n)^m\rceil>1$.
Each building block $G^{i,j}$ of $G$ has at most $n+88n^3$ vertices, without counting $v_{\oplus}^{i,j}$ and $v_{\ominus}^{i,j}$,
because each variable contributes one vertex and each clause 11 vertices. To each building block in layer $i$,
$1\leq i\leq h-1$, at most $56n^3$ building blocks are attached, because each clause contributes 7
building blocks. Let $b(i)$ be the number of building blocks of $G$ in layer $i$. Then, $b(1)=1$ and
$b(i)\leq 56n^3b(i-1)$, where $2\leq i\leq h$. Therefore, since $G$ has $h$ layers, $G$ has at most
$\frac{(56n^3)^h-1}{56n^3-1}$ building blocks. Hence, $G$ has at most
$(n+88n^3)\frac{(56n^3)^h-1}{56n^3-1}+5$ vertices, because each block contributes at most
$n+88n^3$ vertices; plus the 5 vertices of the starting path. Increasing this quantity in order to make
it simpler and substituting $h$ with $\lceil(\log n)^m\rceil$, it turns out that $G$ has at most $2^{12(\log n)^{m+1}}$ vertices.

By lemma~\ref{l:7spanner}, $G$ admits a $v$-concentrated tree 7-spanner; so, $G$ admits a $v$-concentrated
tree $t$-spanner as well. Therefore, algorithm {\tt get\_spanner} on input ($G, v$) returns a $v$-concentrated
tree $tf(G)$-spanner $T$ of $G$. Assume, towards a contradiction, that algorithm {\tt 3-SAT} on input $\phi$
does not return {\tt YES}. Then, first, $|V(G)|>|V(H')|$, because otherwise the exhaustive search would have found
a truth assignment that satisfies $\phi$. Second, for every building block of $G$ truth assignment $a$ defined upon
this building block and $T$ does not satisfy $\phi$.

Here, $H'$ corresponds to $\epsilon=\frac{4}{t12^{\frac{m}{m+1}}}$.
So, $f(G)< \frac{4}{t12^{\frac{m}{m+1}}}(\log n'(G))^{\frac{m}{m+1}}$, because $|V(G)|>|V(H')|$. But
$n'(G)\leq 2^{12(\log n)^{m+1}}$; therefore, $tf(G)< 4(\log n)^m$. Hence, $T$ is a $4h$-spanner of $G$.

For every $i$, $1\leq i\leq h$, there is a building block $G^{i,j}$ of $G$ in layer $i$, such that
$d_T(v_{\oplus}^{i,j},v_{\ominus}^{i,j})=4i$. This is proved by induction on $i$. For $i=1$ there is only one block in layer 1 and
$d_T(v_{\oplus}^{1,1},v_{\ominus}^{1,1})=4$, because of path $P:$ $v_{\oplus}^{1,1}=q_1,p_1,v,p_2,q_2=v_{\ominus}^{1,1}$. Note that
$T$ is $v$-concentrated; so, $P$ is a sub-path of $T$.

For $i>1$, consider layer $i-1$. Then, by induction hypothesis, there is a building block
$G^{i-1,s}$, such that $d_T(v_{\oplus}^{i-1,s},v_{\ominus}^{i-1,s})=4(i-1)$.
Let $a$ be the truth assignment defined by algorithm
{\tt 3-SAT} upon $G^{i-1,s}$ and $T$. Since $a$ does not satisfy $\phi$ there is a clause $c$ in $\phi$ which
is not true through $a$.

Let $X_c$ be the set of the 3 variables that appear in clause $c$. Let $X=\bigcup_{x\in X_c}\{x^{i-1,s},x_c^{i-1,s}\}$.
Since $T$ is a $v$-concentrated spanning tree of $G$, each vertex in $X$ must be adjacent to exactly one of
$v_{\oplus}^{i-1,s}$ or $v_{\ominus}^{i-1,s}$ in $T$.
This holds because, first, $v_{\oplus}^{i-1,s}$ and $v_{\ominus}^{i-1,s}$ are the only
$G$ neighbors of vertices in $X$ that are at $G$ distance
at most\footnote{In a $v$-concentrated spanning tree $T$ of a graph $G$, any vertex at $G$ distance $d$ ($d>0$) from
$v$ must be adjacent in $T$ to a vertex at $G$ distance at most $d$ from $v$.} $d_G(X,v)$ from $v$ (there is no edge
of $G$ between vertices in $X$ and all vertices in $X$ are at the same distance from $v$).
Second, vertices within $G$ distance $d_G(X,v)-1$ from $v$ (here, $v_{\oplus}^{i-1,s}$ and $v_{\ominus}^{i-1,s}$
are at $G$ distance $d_G(X,v)-1$ from $v$) induce a connected sub graph of $T$;
so, a vertex in $X$ cannot be adjacent in $T$ to two vertices at $G$ distance $d_G(X,v)-1$ from $v$.

Let $Q=\bigcup_{r=1}^{r=8}\{q_{r,c}^{i-1,s}\}$. Again, vertices in $X$ are the only
$G$ neighbors of vertices in $Q$ that are at $G$ distance at most $d_G(Q,v)$ from $v$ (graph $G[Q]$ has no edges and
all vertices in $Q$ are at the same distance from $v$).
Also, vertices within $G$ distance $d_G(Q,v)-1$ from $v$ induce a connected sub graph of $T$. So, since
$T$ is a $v$-concentrated spanning tree of $G$, each vertex in $Q$
is adjacent in $T$ to exactly one vertex in $X$.

The $G$ edges between $X$ and $Q$ are these determined
by matrix $M$. But every sub-matrix of $M$ consisting of whole rows of $M$ must contain two complementary to each other
rows of $M$ in order the sub-matrix to have a 1 in each column. Therefore, there is a variable $y$ in $X_c$, such that there is
a vertex in $Q$ adjacent to $y^{i-1,s}$ in $T$ and another vertex in $Q$ adjacent to $y_c^{i-1,s}$ in $T$.
Here, truth assignment $a$ does not make $c$ true; so, if  $y^{i-1,s}$ is adjacent in $T$ to one of
$v_{\oplus}^{i-1,s}$ or $v_{\ominus}^{i-1,s}$, then
$y_c^{i-1,s}$ must be adjacent in $T$ to the other. Therefore, there is a vertex in $Q$ which is at $T$ distance
2 from $v_{\oplus}^{i-1,s}$ and another vertex in $Q$ which is at $T$ distance 2 from $v_{\ominus}^{i-1,s}$. But each vertex
in $Q$ is at $T$ distance 2 from $v_{\oplus}^{i-1,s}$ or $v_{\ominus}^{i-1,s}$; so, there are two consecutive vertices in $Q$,
$q_{r_0,c}^{i-1,s}$ and $q_{r_0+1,c}^{i-1,s}$ say (where $r_0$ is some integer from 1 to 7),
such that $q_{r_0,c}^{i-1,s}$ is at $T$ distance 2 from
one of  $v_{\oplus}^{i-1,s}$ or $v_{\ominus}^{i-1,s}$ and $q_{r_0+1,c}^{i-1,s}$ is at $T$ distance 2 from the other. But
$d_T(v_{\oplus}^{i-1,s},v_{\ominus}^{i-1,s})=4(i-1)$; so, $d_T(q_{r_0,c}^{i-1,s},q_{r_0+1,c}^{i-1,s})=4i$. To pair $q_{r_0,c}^{i-1,s}$ and $q_{r_0+1,c}^{i-1,s}$ is attached a building block of layer $i$, say building block $G^{i,j}$. So,
$d_T(v_{\oplus}^{i,j},v_{\ominus}^{i,j})=4i$ and the induction step holds.

Then, let $G^{h,j}$ be a building block of layer $h$ such that $d_T(v_{\oplus}^{h,j},v_{\ominus}^{h,j})=4h$. Let $x$ be a variable
of $\phi$. Then, $x^{h,j}$ is adjacent\footnote{Note that just one vertex of $G^{h,j}$ (other than $v_{\oplus}^{h,j}$ or
$v_{\ominus}^{h,j}$) is needed. So, in algorithm {\tt reduction} (figure~\ref{f:reduction})
the last layer (layer $h$) of $G$ can be filled instead with
graphs much smaller than building blocks (a path of length 2 suffices) but this does not decrease the number of vertices
of $G$ dramatically.} in $G$ to both of $v_{\oplus}^{h,j}$ and $v_{\ominus}^{h,j}$. But $x^{h,j}$ is adjacent in $T$ to only one
of $v_{\oplus}^{h,j}$ or $v_{\ominus}^{h,j}$, because $T$ is a $v$-concentrated spanning tree of $G$.
Therefore, $x^{h,j}$ is at $T$ distance $4h+1$ from
one of its $G$ neighbors $v_{\oplus}^{h,j}$ or $v_{\ominus}^{h,j}$, which is a contradiction, because $T$ is a $4h$-spanner of $G$.

It remains to check the time complexity of algorithm {\tt 3-SAT} based on the number of variables of input. Construction
of graph $G$ takes $2^{O((\log n)^{m+1})}$ time, because there are at most $2^{12(\log n)^{m+1}}$ vertices in $G$
(and even fewer building blocks in $G$) and each building block of $G$ is constructed efficiently. The exhaustive search takes
place only for small values of $n$. Algorithm {\tt  get\_spanner} is assumed to be efficient but, because of its big input, its call
takes $2^{O((\log n)^{m+1})}$ time. Finally, each building block of $G$ is examined once and each such examination is done
efficiently. Therefore, the {\tt for} loop over building blocks of $G$ takes $2^{O((\log n)^{m+1})}$ time.\myqed

\section{Notes}
The tree 7-spanner returned by algorithm {\tt tree\_7-spanner} in figure~\ref{f:7spanner} is not only $v$-concentrated
but also a breadth first search tree of $G$ starting from $v$, as pointed out in the proof of lemma~\ref{l:7spanner}.
Moreover, restricting algorithm {\tt get\_spanner} to return a breadth first search tree of $G$ starting from $v$, does
not affect the proof of theorem~\ref{t:inapprox}. Therefore, the hardness of approximation described by
theorem~\ref{t:inapprox} also holds for breadth first search trees starting from $v$, which is an even smaller
than $v$-concentrated family of spanning trees.

A few, unrelated to each other, notes follow.
First, this approach does not lead to hardness of approximating tree spanners via general spanning trees; good tree
spanners are not usually breadth first search trees. Second, the result of this article holds for stretch factor $t$ greater
or equal to 7; its an open problem to find low factor approximate tree $t$-spanners for $3\leq t\leq 6$. Third,
function $f=(\log n)^{\frac{\log\log\log n}{\log\log\log n+1}}$ is $o(\log n)$ but there is no $m$, such that
$f$ is $o((\log n)^{\frac{m}{m+1}})$.

\bibliographystyle{plain}
\bibliography{tspanners}

\begin{thebibliography}{10}

\bibitem{Arikati96}
Srinivasa Arikati, Danny~Z. Chen, L.~Paul Chew, Gautam Das, Michiel Smid, and
  Christos~D. Zaroliagis.
\newblock Planar spanners and approximate shortest path queries among obstacles
  in the plane.
\newblock In {\em Algorithms---ESA '96 (Barcelona)}, pages 514--528. Springer,
  Berlin, 1996.

\bibitem{Awerbuch85}
Baruch Awerbuch.
\newblock Complexity of network synchronization.
\newblock {\em Journal of the ACM}, 32(4):804--823, October 1985.

\bibitem{Bandelt86}
Hans-J{\"u}rgen Bandelt and Andreas Dress.
\newblock Reconstructing the shape of a tree from observed dissimilarity data.
\newblock {\em Adv. in Appl. Math.}, 7(3):309--343, 1986.

\bibitem{Bondy89}
J.~A. Bondy.
\newblock Trigraphs.
\newblock {\em Discrete Mathematics}, 75:69--79, 1989.

\bibitem{Brandstadtchordal}
Andreas Brandst{\"{a}}dt, Feodor~F. Dragan, Ho{\`{a}}ng{-}Oanh Le, and Van~Bang
  Le.
\newblock Tree spanners on chordal graphs: complexity and algorithms.
\newblock {\em Theor. Comput. Sci.}, 310(1-3):329--354, 2004.

\bibitem{CaiCor95a}
Leizhen Cai and Derek~G. Corneil.
\newblock Tree spanners.
\newblock {\em SIAM J. of Discrete Mathematics}, 8(3):359--378, 1995.

\bibitem{Chew89}
L.~Paul Chew.
\newblock There are planar graphs almost as good as the complete graph.
\newblock {\em J. Comput. System Sci.}, 39(2):205--219, 1989.
\newblock Computational geometry.

\bibitem{DraganK}
Feodor~F. Dragan and Ekkehard K{\"{o}}hler.
\newblock An approximation algorithm for the tree t-spanner problem on
  unweighted graphs via generalized chordal graphs.
\newblock {\em Algorithmica}, 69(4):884--905, 2014.

\bibitem{pelegemek}
Yuval Emek and David Peleg.
\newblock Approximating minimum max-stretch spanning trees on unweighted
  graphs.
\newblock {\em {SIAM} J. Comput.}, 38(5):1761--1781, 2008.

\bibitem{Fakcharoenphol}
Jittat Fakcharoenphol, Satish Rao, and Kunal Talwar.
\newblock A tight bound on approximating arbitrary metrics by tree metrics.
\newblock {\em Journal of Computer and System Sciences}, 69(3):485 -- 497,
  2004.
\newblock Special Issue on \{STOC\} 2003.

\bibitem{Fekete01}
S{\'a}ndor~P. Fekete and Jana Kremer.
\newblock Tree spanners in planar graphs.
\newblock {\em Discrete Appl. Math.}, 108(1-2):85--103, 2001.
\newblock International Workshop on Graph-Theoretic Concepts in Computer
  Science (Smolenice Castle, 1998).

\bibitem{Fomin}
Fedor~V. Fomin, Petr~A. Golovach, and Erik~Jan van Leeuwen.
\newblock Spanners of bounded degree graphs.
\newblock {\em Inf. Process. Lett.}, 111(3):142--144, 2011.

\bibitem{Narasimhanbook}
Giri Narasimhan and Michiel Smid.
\newblock {\em Geometric Spanner Networks}.
\newblock Cambridge University Press, New York, NY, USA, 2007.

\bibitem{PhDthesis}
Ioannis Papoutsakis.
\newblock {\em Tree Spanners of simple graphs}.
\newblock PhD thesis, Department of Computer Science, University of Toronto,
  2013.
\newblock (Available at university T-space).

\bibitem{Treespannersofsmalldiameter}
Ioannis Papoutsakis.
\newblock Tree spanners of small diameter.
\newblock {\em CoRR}, abs/1503.06063, 2014.

\bibitem{Treespannersofboundeddegree}
Ioannis Papoutsakis.
\newblock Tree spanners of bounded degree graphs.
\newblock {\em CoRR}, abs/1503.06822, 2015.

\bibitem{PelUpf88}
D.~Peleg and E.~Upfal.
\newblock A tradeoff between space and efficiency for routing tables.
\newblock In {\em STOC: ACM Symposium on Theory of Computing (STOC)}, 1988.

\bibitem{PelegReshef}
David Peleg and Eilon Reshef.
\newblock Low complexity variants of the arrow distributed directory.
\newblock {\em Journal of Computer and System Sciences}, 63(3):474 -- 485,
  2001.

\bibitem{Peleg89}
David Peleg and Jeffrey~D. Ullman.
\newblock An optimal synchronizer for the hypercube.
\newblock {\em SIAM J. Comput.}, 18(4):740--747, 1989.

\bibitem{Pettie-Low-Dist-Span}
S.~Pettie.
\newblock Low distortion spanners.
\newblock {\em ACM Transactions on Algorithms}, 6(1), 2009.

\bibitem{Rabinov98}
Y.~Rabinovich and R.~Raz.
\newblock Lower bounds on the distortion of embedding finite metric spaces in
  graphs.
\newblock {\em Discrete Comput. Geom.}, 19(1):79--94, 1998.

\bibitem{Spielman}
Daniel~A. Spielman and Shang-Hua Teng.
\newblock Nearly-linear time algorithms for graph partitioning, graph
  sparsification, and solving linear systems.
\newblock In {\em Proceedings of the Thirty-sixth Annual ACM Symposium on
  Theory of Computing}, STOC '04, pages 81--90, New York, NY, USA, 2004. ACM.

\bibitem{West}
D.~B. West.
\newblock {\em Introduction to Graph Theory}.
\newblock {Prentice Hall, Inc.}, 1996.

\end{thebibliography}
\end{document}